\documentstyle[prl,aps,epsfig]{revtex}
\newcommand{\qvec}{{\bf q}}

\newcommand{\rvec}{{\bf r}}
\newcommand{\rpvec}{{\bf r}'}
\newcommand{\xvec}{{\bf x}}

\newcommand{\beq}{\begin{equation}}
\newcommand{\eeq}{\end{equation}}
\newcommand{\beqa}{\begin{eqnarray}}
\newcommand{\eeqa}{\end{eqnarray}} 

\begin{document}
\draft
\title{Free energy of the Fr\"ohlich polaron in two and three dimensions.} 

\author{John T. Titantah, Carlo Pierleoni and Sergio Ciuchi}

\address{
INFM, UdR L'Aquila and Dipartimento 
di Fisica, Universit\`a de L'Aquila,\\
via Vetoio, I-67100 Coppito-L'Aquila, Italy.}

\date{\today}
\maketitle

\begin{abstract}
We present a novel Path Integral Monte Carlo scheme to solve the Fr\"ohlich polaron model.
At intermediate and strong electron-phonon coupling, the polaron self-trapping is properly taken into 
account at the level of an effective action obtained by a preaveraging procedure with a retarded trial action.
We compute the free energy at several couplings and temperatures in three and 
two dimensions. Our results show that the accuracy of the Feynman variational upper bound for the free energy 
is always better than $5\%$ although the thermodynamics derived from it is not correct.
Our estimates of the ground state energies demonstrate that the second cumulant correction to 
the variational upper bound predicts the self energy to better than $1\%$ at intermediate and strong coupling. 
\end{abstract}

\section{Introduction}
The quasiparticle formed by an electron added to a polar crystal and the lattice deformation 
associated to it is called a ``polaron'' after Landau\cite{Landau33}.
For non localized electrons in polar insulators and semiconductors, Fr\"ohlich et al.
\cite{FrohlichPelzerZienau50}
introduced a model for a single electron interacting with dispersionless longitudinal optical phonons.
Few years later, within the path integral formalism, Feynman proposed a very powerful variational 
procedure to compute the ground state energy and the effective mass of the model which provide
accurate results at all couplings \cite{Feynman55}.
~Extension of the Feynman technique to compute
properties at finite temperature appeared few years later by Osaka \cite{Osaka}, 
and more recently by Castrigiano et al. \cite{cas83-84} (see also Ref.~\onlinecite{KhandekarLawande}). 
Recently the polaron problem has gained new interest
as it could play a role in explaining the properties of the high $T_c$ superconducting materials\cite{Stripes98}.

After a first Monte Carlo attempt by Becker et al. \cite{BeckerGerlachSchliffke}, a true numerical test of the 
Feynman solution was provided by Alexandrou et al. \cite{AlexandrouRosenfelder}.
Applying an original extension of the Fourier Path Integral Monte Carlo (PIMC) method to retarded actions,
and special importance sampling technique, they provided exact (in the Monte Carlo sense) results
for the ground state energy and the effective mass of the Fr\"ohlich polaron in three dimensions.

Recently the same model has been studied 
with a diagrammatic Monte Carlo (MC) technique.
Attention has been focused on the ground state energy \cite{prokfev98},
effective mass, phonon distribution and on the excitation spectrum at zero
temperature \cite{prokfev00}.
Also the effective mass for a lattice version of the Fr\"ohlich polaron in two 
dimensions has been obtained \cite{Kornilovitch98}.

In this letter we extend the numerical work on the Fr\"ohlich polaron to finite temperature and 
we provide results for the free energy in three and two dimensions. We focus on the intermediate and 
strong coupling regime which was only partially covered by the previous ground state MC investigations.
Extrapolation of the free energy to zero temperature provides 
the ground state energy. The result of our analysis is that the Feynman variational method (FVM) at finite 
temperature provides estimates for the free energy within few percent at all couplings and dimensions, 
as it was already demonstrated for the ground state energy\cite{AlexandrouRosenfelder}. 
However, due to its variational nature,
FVM does not provide a consistent solution of the problem as thermodynamic derivatives of the
free energy are not given correctly. 
In particular the entropy of the variational solution does not vanish with temperature as it should
according to the third principle of thermodynamics. 
This is not just an academic issue since 
in the applications where the model could be invoked thermal effects are likely to be relevant. Indeed the 
phonon frequency in real materials is of the order of tenths of $meV$. 

We have formulated the problem in the framework of the discretized time PIMC method \cite{Ceperley-rmp}.
The high temperature density matrix has been obtained through an extension of the preaveraging 
technique \cite{DollFreemanBeck} (or cumulant method\cite{Ceperley-rmp}) to retarded actions. 
This procedure regularizes the attractive coulomb-like divergence of the bare action at short distance (see below), 
and makes well defined and stable the discretized model\cite{CiuchiLorenzanaPierleoni}.
Furthermore, in order to cope with the polaron self-trapping phenomena,
we propose a suitable retarded trial action (RTA) around which the 
preaveraging procedure is more effective than the standard preaveraging with the free particle 
trial action (FPTA). 
At intermediate and strong coupling and at low temperature, the RTA improves the convergence 
with the number of time slices $M$ compared to the FPTA. 
Moreover, at given coupling, it removes a cross-over in the convergence of thermodynamic quantities 
with the number of slices and 
allows to obtain reliable extrapolations ($M \rightarrow \infty$) 
from simulations with limited number of slices only.
The methodological aspects of the present work are essential to attack the problem of the simulation of 
many interacting large polarons 
a system which has received some attention\cite{Lemmens,Mahan90,Cataudella,Fratini} 
as it could be relevant to describe the unscreened long range interactions in the $CuO$ planes of the 
cuprates high $T_c$ superconductors\cite{Calvani}.
It is known that the free energy differences between different phases of the electron gas are quite 
small\cite{JonesCeperley96}. The characterization of the electron-phonon effects on these phases requires 
an accurate method to compute the free energy for the many polaron system.
The method presented here is similar to the one used by Alexandrou and Rosenfelder but in real rather than in 
Fourier space\cite{AlexandrouRosenfelder}.

We start from the effective action of the Fr\"ohlich polaron,
after the phonons have been eliminated with the usual path integral
technique\cite{Feynman}. In polaronic units, $\hbar\omega_{LO}=1, \sqrt{\hbar/(m\omega_{LO})}=1$
($\omega_{LO}$ is the phonon frequency and $m$ the electron band mass), and 
discretizing the imaginary time interval $\beta$ in $M$ intervals of width $\tau$ it is
\begin{eqnarray}
  \label{eq:eff-act}
  &S=&S_0+S_{e-ph}=\frac{1}{2}\sum_{i=1}^M\int_{(i-1)\tau}^{i\tau} dt_1 \dot{\qvec}^2 \\
&-&\frac{\alpha}{2\sqrt{2}} \sum_{i=1}^M \sum_{j=1}^M
\int_{(i-1)\tau}^{i\tau} dt_1 \int_{(j-1)\tau}^{j\tau} dt_2 
\frac{ D(t_1-t_2)}{|\qvec(t_1)-\qvec(t_2)|}\nonumber
\end{eqnarray}
Here $\alpha$ is the coupling constant, $\qvec(t)$ is the electron position at imaginary time $t$ and
$D(t,\beta)=[e^t + e^{\beta-t}]/[e^{\beta}-1]$
is the phonon propagator. In the present work we consider the case of isotropic band mass ($D=3$) and 
the case in which the band mass in one direction is infinitely large and the motion of the 
electron occurs on a plane ($D=2$)\cite{Xiaoguang}.
The density matrix for the electron to go from $\rvec$ to $\rpvec$ in a time $\beta$ is
\beqa
\label{eq:denmat}
\rho(\rvec,\rpvec|\beta)&=&   
\int \Pi_{j=1}^{M-1} d\qvec_j \int {\cal D\xvec}_j e^{-(S-S_T)}~e^{-S_T} \\ \nonumber
&=&\int \Pi_{j=1}^{M-1} d\qvec_j \left\{ \left<e^{-(S-S_T)}\right>_{T_j} \int {\cal D\xvec}_j~e^{-S_T} \right\}
\eeqa
where $\qvec_j=\qvec(j\tau)$, 
$\qvec_0=\qvec(0)=\rvec,~\qvec_M=\qvec(\beta)=\rpvec$, 
$\xvec_j(t)$ represent the paths of the $j-th$ time slice (from $\qvec_{j-1}$
to $\qvec_j$) 
and we have introduced a retarded trial action quadratic 
in the path coordinates 
\beqa
\label{eq:trialact}
S_T&=&S_0\\ \nonumber
   &+&\frac{C}{2}\sum_{i=1}^M \sum_{j=1}^M D_{ij}^w
\int_{(i-1)\tau}^{i\tau} dt_1 \int_{(j-1)\tau}^{j\tau} dt_2 
{|\qvec(t_1)-\qvec(t_2)|^2}
\eeqa
where $D_{ij}^w=D(w\tau|i-j|,w\beta)$ and $C$ and $w$ are variational parameters. 
In eq. (\ref{eq:denmat}) $\left<...\right>_{T_j}$ represents an average over paths belonging to the 
$j-th$ time slice weighted by the trial action $S_T$. 
Using the Fourier representation of each time interval, and taking advantage of the 
gaussian character of the Fourier amplitudes which arises from the quadratic trial action (\ref{eq:trialact}),
we can analytically perform the preaveraging procedure\cite{DollFreemanBeck,TitantahCiuchiPierleoni} 
to obtain an effective action between 
any pair of slices in terms of the end points of the slices only 
\beqa
\label{eq:preave}
\rho(\rvec,\rpvec|\beta) \geq \rho_c(\rvec,\rpvec|\beta)=&\\ \nonumber
\int \Pi_{i=1}^{M-1} d\qvec_i 
\left\{ e^{-<S-S_T>_{T_i}} \int {\cal D\xvec}_i~e^{-S_T} \right\} = \\ \nonumber
\left(\frac{1}{2\pi\tau}\right)^{Md/2}
\int \Pi_{i=1}^{M-1} d\qvec_i~e^{-(S_0(\{\qvec_i\}) + S_{eff}(\{\qvec_i\}))}
\eeqa
For $C=0$ the
standard free particle trial action local in imaginary time is implied and we get
\beqa
S_0&=&\frac{1}{2\tau}\sum_{i=1}^M |\qvec_i-\qvec_{i-1}|^2 \\
S_{eff}&=&-\frac{\alpha\tau^2}{2\sqrt 2}\sum_{i=1}^M \sum_{j=1}^M
\int_{-1/2}^{1/2} du \int_{-1/2}^{1/2} dv
\frac{D(\tau|i-j+u-v|)}{|{\bf L}_{ij}(u,v)|}
~G\left(\frac{|{\bf L}_{ij}(u,v)|}{\Sigma_{ij}(u,v)}\right)
\label{eq:Seff}
\eeqa
where 
\beqa
G(x)&=&erf(x)  \hskip 3.6cm D=3 \\
\label{eq:bessel}
    &=&\sqrt{\frac{\pi}{2}}x~I_o(\frac{x^2}{4})~e^{-x^2/4} \hskip 1.5cm D=2 \\  
{\bf L}_{ij}(u,v)&=& \frac{\qvec_i+\qvec_{i-1}}{2}-\frac{\qvec_j+\qvec_{j-1}}{2}+
u(\qvec_i-\qvec_{i-1})-v(\qvec_j-\qvec_{j-1}) \\
\Sigma_{ij}^2(u,v)&=&\tau[1-2(u^2+v^2)] \hskip 2.1cm i\neq j \\
\Sigma_{ii}^2(u,v)&=&2\tau|u-v|(1-|u-v|)
\eeqa
In eq.(\ref{eq:bessel}), $I_0(x)$ is the modified Bessel function\cite{Abramowitz}.
In the general case $C\neq0$, the expression
for $S_{eff}$ is much longer and we will report it in a 
more extensive publication\cite{TitantahCiuchiPierleoni}. Here we just mention that its derivation is 
long but straightforward.

We have sampled the path configurational space by Metropolis Monte Carlo method\cite{Ceperley-rmp}
with probability distribution proportional to $exp\{-(S_0+S_{eff})\}$. Even at strong coupling,
the simple Levy reconstruction of paths\cite{Ceperley-rmp} was appropriate to efficiently
move through configurational space. Typically between $20$ and $400$ time slices have been used in 
this study. The double integral over $u$ and $v$ in $S_{eff}$ (see eq.(\ref{eq:Seff}))
is performed by Gauss method on the fly\cite{Abramowitz}.

The method outlined above to derive the effective action is variational 
since $F_c=-\beta^{-1}ln[Tr(\rho_c)]\geq F=-\beta^{-1}ln[Tr(\rho)]/\beta$
holds at any finite $\tau$, the equality being valid in the $\tau\rightarrow 0$ limit.
At given $\tau$, optimal trial action can be obtained minimizing 
 the average of the effective action with respect to the parameters $C$ and $w$ in eq.(\ref{eq:trialact}).
This is related to the excess free energy
$F_c^{ex}$ through the thermodynamic integration in the coupling constant $\alpha$
\beq
F_c^{ex}(\alpha)=\frac{1}{\beta}\int_0^{\alpha}d\alpha' \frac{<S_{eff}>_{\alpha'}}{\alpha'}
\label{eq:thermint}
\eeq
We have checked in
few significant cases that the optimal values for $C$ and $w$ are always close to the
Feynman variational values\cite{cas83-84}, which we have used in our simulations.

\begin{figure}[H]
 \centerline{\psfig{file=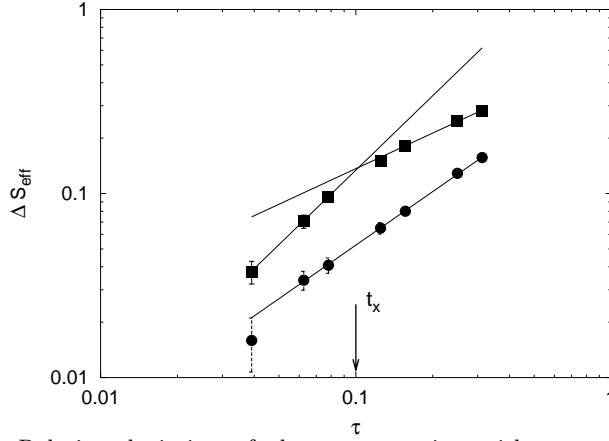,width=6.cm,angle=-90}}
 \caption{$\alpha=9$, $\beta=10$. Relative deviation of the excess action with respect to its extrapolation at 
          $\tau=0$, $\Delta S_{eff}=[<S_{eff}>_{\tau}-<S_{eff}>_0]/<S_{eff}>_0$, versus 
          $\tau$.
	  Comparison between FPTA data (squares) and RTA data (circles).
          Lines are
          power law fits to the data. The crossover $t_x$ discussed in the text is clearly seen in the FPTA data.}
 \label{fig:fig1}
\end{figure}

\begin{figure}[H]
 \centerline{\psfig{file=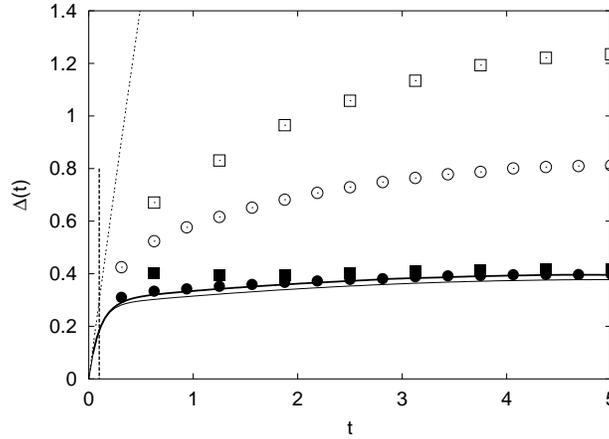,width=6cm,angle=-90}}
 \caption{
$\alpha=9$, $\beta=10$. 
Open (closed) symbols are data obtained with the FPTA (RTA). Data for $M=16$ (squares) and 
$M=32$ (circles) are reported. The thick continuous line is the results of a fully converged
simulation ($M=256$ with the RTA), while the thin continuous line is the prediction of FVM.
The dashed line is the free particle behavior while the dashed vertical line indicates $t_x$.} 
\label{fig:diff}
\end{figure}

We exploited eq.(\ref{eq:thermint}) to obtain the free energy of the polaron.
At given coupling and temperature, the coupling constant integration has been performed 
by Gauss method. We have checked that order four in the Gauss integration
is enough to obtain accurate results even at strong couplings. 
At each Gauss point a series of simulations for increasing number of slices were
necessary to extrapolate the average effective action to $\tau=0$. In figure 
\ref{fig:fig1} we show the typical behavior of the average effective action  
versus $\tau$ at strong coupling and low temperature,
and we compare data obtained with the FPTA ($C=0$) 
and with the RTA ($C\neq 0$).
The RTA improves the convergence at fixed $\tau$ by a factor between two and three and it
removes a crossover in $\tau$ observed in the behavior for the FPTA. It allows
to extract reliable extrapolations from data at large $\tau$ only. This phenomenon can 
be understood comparing the diffusional properties of the paths with the two  
trial actions\cite{TitantahCiuchiPierleoni}.
The imaginary time mean square displacement $\Delta(t)=<|\qvec(t)-\qvec(0)|^2>$ 
is the quantity appropriate to discuss the self-trapping of 
the polaron. Indeed, at strong coupling and at low enough temperature, 
the mean square displacement presents a crossover between a free particle-like regime at short
time $t<t_x$ ($\Delta(t)=t(\beta-t)/\beta$) and a localized regime at larger time $t>t_x$
($\Delta(t)\approx constant$) as it is shown in figure \ref{fig:diff}.
In the same figure we see that the results obtained with the RTA
exhibit the self-trapping phenomena already for a limited number of slices ($\tau>t_x$) and
they converge with $M$ very quickly to the asymptotic behavior.
Conversely, with the FPTA must be $\tau<t_x$ in order to correctly sample the self-trapped polaron.
This explains the crossover in figure \ref{fig:fig1}.

In tables \ref{tab:3D} and \ref{tab:2D} we report the values for the free energy at some selected 
thermodynamic points in three and two dimensions, respectively, and we compare with the results of
the FVM both for the upper and lower bounds\cite{FLB}. 
The Feynman upper bound (FUB) is always close to the PIMC 
data (few percent). At intermediate coupling,
FUB is more accurate in three rather than in two dimensions: the relative deviation of FUB from the PIMC 
data is maximum
for $\alpha=3$ and $D=2$ (between $2.5\%$ and $4.5\%$) which corresponds approximately to 
$\alpha=7$ for $D=3$\cite{Xiaoguang}. Data for $D=3$ and $\alpha=5$ suggest that 
at fixed coupling, the relative deviation
is maximum around $\beta\omega_{LO}=0.1$

\begin{figure}[H]
 \centerline{\psfig{file=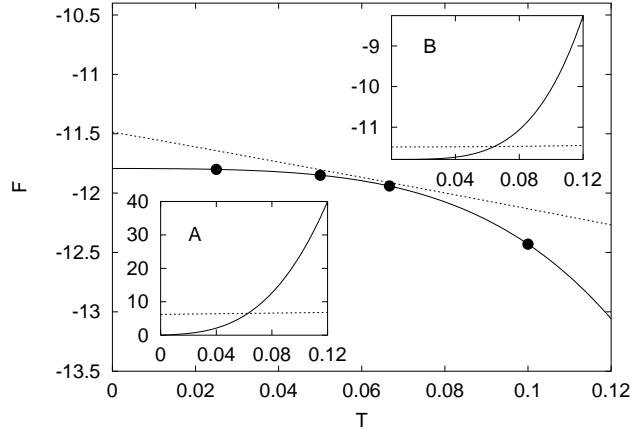,width=6.cm,angle=-90}}
 \caption{Excess free energy versus temperature at $\alpha=9$. Closed circles are PIMC data and the continuous line is a
polynomial fit to the data. The dashed line is the FUB. In the inset A we show the entropy vs temperature as obtained
by deriving with T both the polynomial fit to the PIMC data (continuous line) and the FUB (dashed line).
In the inset B we report the internal energy $F+TS$.}
 \label{fig:FvsT}
\end{figure}

At given $\alpha$, the excess free energy for $T\leq0.1$ is well reproduced by a polynomial function
in terms of the even powers of $T$ only. This is compatible with the third principle of thermodynamics
as the entropy at $T=0$ must vanish. In figure \ref{fig:FvsT} we show the excess free energy versus temperature 
at $\alpha=9, \beta=10$ and $D=3$. We also report in the insets the entropy and the internal energy obtained 
as temperature derivatives of the polynomial function fitted to the PIMC data. Comparison with curves obtained 
from the FUB show the inadequacy of the variational method to provide correct thermodynamic derivatives of the 
free energy. 

We estimate the ground state energy of the polaron by the 
coefficient of the order zero term in the fitting function for the excess free energy. 
The numerical values obtained for 
the considered cases are collected in table \ref{tab:E0}.
~Our data compare well with a previous PIMC estimate\cite{AlexandrouRosenfelder} 
($E_0=-5.537(20)$ at $\alpha=5, D=3$) and 
confirm that predictions based on the calculation of the second cumulant within FVM\cite{MM70} 
are very accurate at any coupling. At $D=2$ no such calculations has been performed.

In conclusion we have developed an efficient scheme to simulate the single Fr\"ohlich polaron model within the 
discretized time Path Integrals Monte Carlo method. The new scheme is based on two key points: i) the extension of 
the preaveraging procedure to retarded actions; ii) the introduction of a suitable quadratic trial action,
similar to the one used by Feynman, around which to perform the preaveraging. 
We have shown that the scheme is able to 
integrate the short (imaginary) time motion of the paths up to times larger than the self-trapping time
of the polaron.
Applying this new method
we have obtained fully converged results for the free energy and the ground state energy of the polaron 
in two and three dimensions at intermediate and strong electron-phonon coupling. 
The present results show that the accuracy of the Feynman upper bound for the free energy 
is always better than $5\%$. 

Our work extends the one of Alexandrou and Rosenfelder\cite{AlexandrouRosenfelder} which however formulated 
the problem in the framework of the Fourier Path Integral Monte Carlo. The advantage of the 
present formulation is that its extension to the many polarons system is straightforward.
The present scheme can be used in connection with the recently proposed 
Restricted Path Integrals MC devised to circumvent the fermion sign problem\cite{Ceperley96}.
Work in this direction is in progress.

\acknowledgments
We acknowledge useful 
discussions and suggestions from D.M. Ceperley, S. Fratini and J. Lorenzana.  
We acknowledge partial support
from the MURST 1999 matching funds program.



\begin{table}
\caption{D=3. Comparison between the PIMC free energy and the upper bound (FUB) 
and the lower bound (FLB) of the Feynman variational method.}
\label{tab:3D}
\begin{tabular}{ddddd}
$\alpha$ & $\beta$ & FLB & PIMC  &FUB\\
\tableline
     1.0 &     1.0 &-1.87001 &-1.838(2) &-1.830 \\
     1.0 &     5.0 & ---     &-1.103(1) &-1.095\\
     1.0 &    10.0 &-1.10077 &-1.053(1) &-1.043\\
\tableline
     3.0 &     1.0 &-5.90912&-5.659(1) &-5.638\\
     3.0 &     5.0 &-3.70067&-3.479(2) &-3.419\\
     3.0 &    10.0 & ---    &-3.251(3) &-3.244\\
\tableline
     5.0 &     1.0 &-10.8646 &-9.827(5) &-9.696\\
     5.0 &     5.0 &-7.16466 &-6.12(1)  &-6.018\\
     5.0 &    10.0 &-6.71199 &-5.800(5) &-5.678\\
     5.0 &    15.0 &-6.49491 &-5.654(3) &-5.590 \\
     5.0 &    20.0 &-6.41828 &-5.593(2) &-5.550\\
     5.0 &    40.0 &-6.37900 &-5.539(2) &-5.493\\ 
\tableline
     9.0 &    10.0 &-17.94248&-12.43(1) &-12.13\\
     9.0 &    15.0 &-17.37586&-11.94(1) &-11.91\\
     9.0 &    20.0 &-17.33489&-11.85(1) &-11.80\\
     9.0 &    40.0 &-17.32615&-11.80(2) &-11.64
\end{tabular}
\end{table}

\begin{table}
\caption{D=2.Comparison between the PIMC free energy and the upper bound (FUB) 
and the lower bound (FLB) of the Feynman variational method.}
\label{tab:2D}
\begin{tabular}{ddddd}\\
$\alpha$ & $\beta$ &  FLB & PIMC     & FUB  \\
\tableline
     1.0 &     10.0&-1.79381     & -1.6940(6) & -1.6778\\
     1.0 &     15.0&-1.76034     & -1.6663(9) & -1.6576\\
     1.0 &     20.0&-1.74893     & -1.6535(7) & -1.6483\\
     1.0 &     40.0&  ---        & -1.6412(9) & -1.6354\\
\tableline
     3.0 &     10.0&-7.66997     & -6.044(5)  & -5.777\\
     3.0 &     15.0&-7.45560     & -5.771(6)  & -5.671\\
     3.0 &     20.0&-7.40605     & -5.716(3)  & -5.620\\
     3.0 &     40.0&  ---        & -5.677(6)  & -5.547
\end{tabular}
\end{table}

\begin{table}
\caption{Polaron ground state energy. Comparison between our 
present estimates (PIMC), variational 
upper and lower bounds (FUP and FLB),
data from ref. [27] 
for $D=3$ and from ref. [23] 
for $D=2$ (PT).}
\label{tab:E0} 
\begin{tabular}{dddddd}\\
$D$ & $\alpha$  &   FLB  &  PIMC     &   FUB  &    PT  \\
\tableline
 3  &   5.0     &-6.360  & -5.523(3) & -5.440 &  -5.52 \\
 3  &   9.0     &-17.292 & -11.794(2)& -11.49 &  -11.7 \\
\tableline
 2  &   1.0     &-1.735  & -1.635(2) & -1.616 &  -1.63477 \\
 2  &   3.0     &-7.402  & -5.670(3) & -5.467 &  -5.28812      
\end{tabular}
\end{table}


\begin{thebibliography}{10}
\bibitem{FLB}
Within Feynman method a lower bound for the 
free energy can be obtained as $ F_T + \beta^{-1} <S - S_T> \leq F$
where $F_T$ is the free energy of the quadratic Feynman action
and 
the average is performed with the true polaron action $S$.
Since  $F_T$ and $-S_T$ are both negative quantities 
the lower bound is obtained when $F_T=S_T=0$ 
and it is simply $<S>/\beta$ independent 
on the trial action.
\end{thebibliography}
\end{document}